\documentclass[aps,prx,twocolumn,superscriptaddress]{revtex4-1}
\usepackage{graphicx,braket,hyperref,amsmath,comment,float}
\hypersetup{
	colorlinks   = true,
	allcolors    = [rgb]{0,0,1.0}
}
\begin{document}

\title{High-pressure insulator-to-metal transition in Sr$_3$Ir$_2$O$_7$ studied by x-ray absorption spectroscopy}

\author{C. Donnerer}
\affiliation{London Centre for Nanotechnology and Department of Physics and Astronomy, University College London, London WC1E 6BT, United Kingdom}

\author{M. Moretti Sala}
\affiliation{ESRF - The European Synchrotron, 71 Avenue des Martyrs, 38000 Grenoble, France}

\author{S. Pascarelli}
\affiliation{ESRF - The European Synchrotron, 71 Avenue des Martyrs, 38000 Grenoble, France}

\author{A. D. Rosa}
\affiliation{ESRF - The European Synchrotron, 71 Avenue des Martyrs, 38000 Grenoble, France}

\author{S.N. Andreev}
\affiliation{Theoretical Physics and Applied Mathematics Department, Ural Federal University, 620002 Ekaterinburg, Russia}

\author{V. V. Mazurenko}
\affiliation{Theoretical Physics and Applied Mathematics Department, Ural Federal University, 620002 Ekaterinburg, Russia}

\author{T. Irifune}
\affiliation{Geodynamics Research Center, Ehime University, 2-5 Bunkyo-cho, Matsuyama 790-8577, Japan}

\author{E. C. Hunter}
\altaffiliation{Present address: Inorganic Chemistry Laboratory, University of Oxford, South Parks Road, Oxford OX1 3QR, UK}
\affiliation{SUPA, School of Physics and Astronomy, and Centre for Science at Extreme Conditions, The University of Edinburgh, Mayfield Road, Edinburgh EH9 3JZ, United Kingdom}

\author{R. S. Perry}
\affiliation{London Centre for Nanotechnology and Department of Physics and Astronomy, University College London, London WC1E 6BT, United Kingdom}

\author{D. F. McMorrow}
\affiliation{London Centre for Nanotechnology and Department of Physics and Astronomy, University College London, London WC1E 6BT, United Kingdom}

\begin{abstract}
High-pressure x-ray absorption spectroscopy was performed at the Ir $L_3$ and $L_2$ absorption edges of Sr$_3$Ir$_2$O$_7$. The branching ratio of white line intensities continuously decreases with pressure, reflecting a reduction in the angular part of the expectation value of the spin-orbit coupling operator, $\braket{{\bf L} \cdot {\bf S}}$. Up to the high-pressure structural transition at 53 GPa, this behavior can be explained within a single-ion model, where pressure increases the strength of the cubic crystal field, which suppresses the spin-orbit induced hybridization of $J_{\text{eff}} = 3/2$ and $e_g$ levels. We observe a further reduction of the branching ratio above the structural transition, which cannot be explained within a single-ion model of spin-orbit coupling and cubic crystal fields. This change in $\braket{{\bf L} \cdot {\bf S}}$ in the high-pressure, metallic phase of Sr$_3$Ir$_2$O$_7$ could arise from non-cubic crystal fields or a bandwidth-driven hybridization of $J_{\text{eff}}=1/2,\,3/2$ states, and suggests that the electronic ground state significantly deviates from the $J_{\text{eff}}=1/2$ limit.
\end{abstract}

\maketitle

\section{Introduction}

Materials that exhibit strong spin-orbit coupling and electronic correlations provide an attractive platform to explore novel electronic and magnetic phenomena \cite{jackeli2009mott, wang2011twisted, wan2011topological, krempa2014correlated, rau2016spin-orbit}. Members of the Ruddlesden-Popper (RP) series of iridates Sr$_{n+1}$Ir$_n$O$_{3n+1}$, where $n$ is the number of consecutive SrIrO$_3$ perovskite layers, have been at the focus of many investigations since the discovery of a spin-orbit induced Mott insulating state in single-layer Sr$_{2}$IrO$_{4}$ ($n=1$) \cite{kim2008novel, kim2009phase}. In these Ir$^{4+}$ ($5d^5$) iridates, the cubic crystal field (CF) $10Dq$ dominates over Hund's coupling $J_H$ ($10Dq > 3 J_H$), resulting in a $t_{2g}^5$ configuration. Subsequently, the spin-orbit coupling (SOC) splits the $t_{2g}$ manifold into a $J_{\text{eff}}=1/2$ doublet and a $J_{\text{eff}}=3/2$ quartet (where $J_{\text{eff}} = \left| -{\bf L} + {\bf S}\right|$). On-site Coulomb interactions can then open an energy gap in the half-filled $J_{\text{eff}}=1/2$ band, giving rise to a $J_{\text{eff}}=1/2$ Mott insulator. As the dimensionality increases with the number of perovskite layers $n$, the insulating gap closes in the RP iridates, eventually reaching a metallic state in SrIrO$_3$ ($n=\infty$) \cite{moon2008dimensionality, zhang2013effective}. The insulating bilayer Sr$_3$Ir$_2$O$_7$ ($n=2$) is in close proximity to this insulator-metal boundary, such that it can be metallized by application of high-pressure \cite{ding2016pressure}, which offers insights into the nature of the Mott transition in the strong spin-orbit coupling limit.

At ambient pressure, the crystal structure of Sr$_3$Ir$_2$O$_7$ can be approximated by a tetragonal model (space group $I4/mmm$), where the in-plane IrO$_6$ rotations ($\sim 12^\circ$) are treated as disordered \cite{subramanian1994single}. Other studies have claimed that the octahedral rotations are correlated, resulting in an orthorhombic, twinned structure (space group $Bbcb$) \cite{cao2002anomalous, matsuhata2004crystal}, as well as finding an out-of-plane octahedral tilt ($\sim 0.2^\circ$), which lowers the symmetry to monoclinic (space group $C2/c$) \cite{hogan2016structural}. Early high-pressure resistivity and x-ray diffraction (XRD) studies gave a somewhat conflicting account of the evolution of electronic and structural properties of Sr$_3$Ir$_2$O$_7$ \cite{li2013tuning, zhao2014pressure, zocco2014persistent}. Recent high-pressure XRD, resistance, and resonant inelastic x-ray scattering (RIXS) measurements have converged on a clearer picture \cite{donnerer2016pressure, ding2016pressure}: Given a modest charge-gap of $\sim$ 100 meV \cite{okada2013imaging, wojek2012the, wang2013dimensionality, king2013spectroscopic, park2014phonon}, it was found surprising that no metallization occurred in Sr$_3$Ir$_2$O$_7$ up to 50 GPa. This could be understood by an increase in octahedral rotation, which alleviates the pressure-induced bandwidth broadening \cite{donnerer2016pressure, solovyev2015validity}. At 53 GPa, XRD showed a reversible first-order transition to a high-pressure structure (space group $C2$), which adopts a modified stacking sequence of the perovskite bilayers \cite{donnerer2016pressure}. At the same pressure, or in close proximity, resistance measurements found an insulator-to-metal transition (IMT) \cite{ding2016pressure}. It is thus likely that structural and electronic transitions are coupled. Intriguingly, resistance measurements showed conductance in the $ab$ plane, but an insulating behavior along the $c$ axis, leading to the proposal that the high-pressure phase of Sr$_3$Ir$_2$O$_7$ is a ``confined metal" \cite{ding2016pressure}.

However, the structural, electronic, and magnetic properties of the high-pressure phase of Sr$_3$Ir$_2$O$_7$ remain poorly understood. While RIXS measurements have been able to follow crystal field excitations up to 65 GPa, low-energy transfer features become less well-defined above 20 GPa. It is hence difficult to quantify subtle changes in the electronic state, such as the presence of non-cubic crystal fields, and to what extent a $J_{\text{eff}} = 1/2$ description remains valid.

Here, we investigate the role of spin-orbit coupling, crystal fields and electronic bandwidth in Sr$_3$Ir$_2$O$_7$ at high pressure via x-ray absorption spectroscopy (XAS) at the Ir $L_3$ and $L_2$ absorption edges. The branching ratio (BR) of white line intensities ($I_{L_3}/I_{L_2}$) offers a unique insight into the character of the electronic ground state, as it is directly proportional to the angular part of the expectation value of the spin-orbit coupling operator, $\braket{{\bf L} \cdot {\bf S}}$ \cite{vanderLaan1988local, thole1988linear}. Up to 50 GPa, we find a gradual reduction of the BR with pressure, from $I_{L_3}/I_{L_2} \sim 5.5$ to $\sim 4$. This can be understood by an increased cubic CF splitting at high pressures, which suppresses the spin-orbit induced mixing of $J_{\text{eff}}=3/2$ and $e_g$ states; reducing the $\braket{{\bf L} \cdot {\bf S}}$ contribution from holes in the $e_g$ states. Above the structural transition at 53 GPa, a further reduction in the BR occurs, which cannot be explained by considering cubic crystal fields and spin-orbit coupling on a single-ion level. We propose that the high-pressure structural transition could result in non-cubic crystal fields or an increased electronic bandwidth, both of which would lead to hybridization of $J_{\text{eff}}=1/2$ and $3/2$ states. This indicates a departure from the $J_{\text{eff}}=1/2$ model in the high-pressure phase of Sr$_3$Ir$_2$O$_7$.

\section{Experimental method}

High-pressure energy-dispersive XAS experiments were performed at beamline ID24 of the European Synchrotron Radiation Facility (ESRF) \cite{pascarelli2016the}. We conducted independent experiments at the Ir $L_3$ and $L_2$ absorption edges. Single crystals of Sr$_3$Ir$_2$O$_7$ were flux-grown as described in Ref. \onlinecite{boseggia2012antiferromagnetic}. Symmetrical diamond anvil cells (DACs) fitted with polycrystalline diamond anvils were used. Single crystals of Sr$_3$Ir$_2$O$_7$ were ground into a powder, pressed into a pellet, and then loaded into the DACs. Pressure was measured \emph{in situ} using ruby fluorescence. Neon was used as the pressure transmitting medium. All data were taken at room temperature.

\section{Results and Discussion}

\begin{figure}[t]
\includegraphics[width=\linewidth]{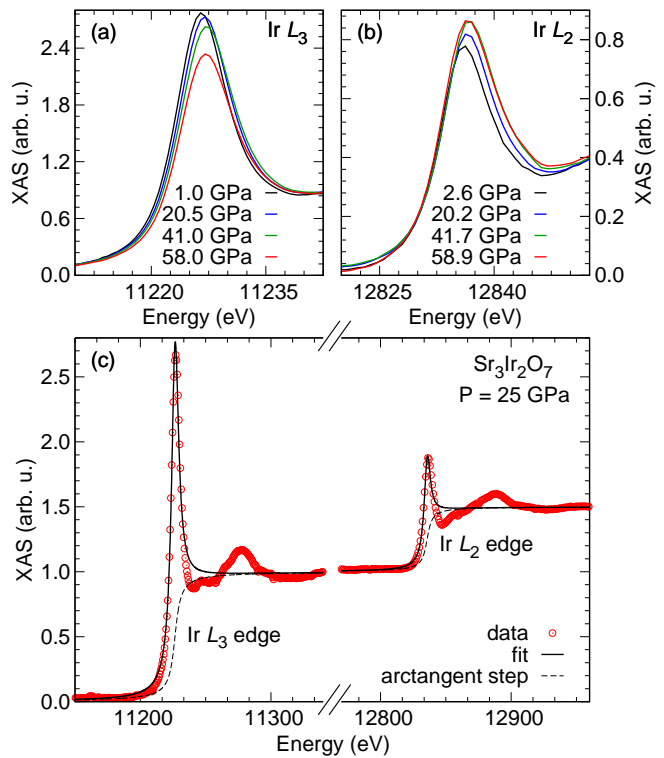}
\caption{High-pressure XAS data of Sr$_3$Ir$_2$O$_7$. Pressure evolution of XAS at the Ir (a) $L_3$ and (b) $L_2$ absorption edges. (c) Representative fitted XAS at 25 GPa at the Ir $L_3$ and $L_2$ absorption edges of Sr$_3$Ir$_2$O$_7$. The red circles are the normalized data points, the black solid line is a fit of arctangent and a Lorentzian functions, as described in the text.}
\label{fig1}
\end{figure}

Figure \ref{fig1} shows representative high-pressure x-ray absorption spectra at the Ir $L_3$ and $L_2$ edges of Sr$_3$Ir$_2$O$_7$. The data were normalized to absorption steps of 1 and 1/2 for $L_3$ and $L_2$ edges, respectively.

The branching ratio of integrated white line intensities, $\text{BR} = I_{L_3}/I_{L_2}$, can be directly related to the angular part of the expectation value of the spin-orbit coupling, via BR $ = (2+r)/(1-r)$, where $r = \braket{{\bf L} \cdot {\bf S}}/n_h$ and $n_h$ is the number of holes in the $5d$ manifold \cite{thole1988linear, vanderLaan1988local}. Without performing any data analysis, it is clear that the observed BR exceeds the statistical value of 2 (obtained by setting $\braket{{\bf L} \cdot {\bf S}} = 0$; we define $\hbar = 1$ throughout the paper), which shows that the spin-orbit coupling plays a significant role in the electronic ground state. Near ambient pressure, we estimate a BR of $\sim 5.5$, which, by using $n_h=5$ for Ir$^{4+}$ ($5d^5$), yields $\braket{{\bf L} \cdot {\bf S}} \sim 2.7$. This SOC expectation value is characteristic for Ir$^{4+}$ in a cubic crystal field \cite{laguna-marco2010orbital, clancy2012spin, haskel2012pressure, laguna-marco2015electronic}.

\begin{figure}[t]
\includegraphics[width=\linewidth]{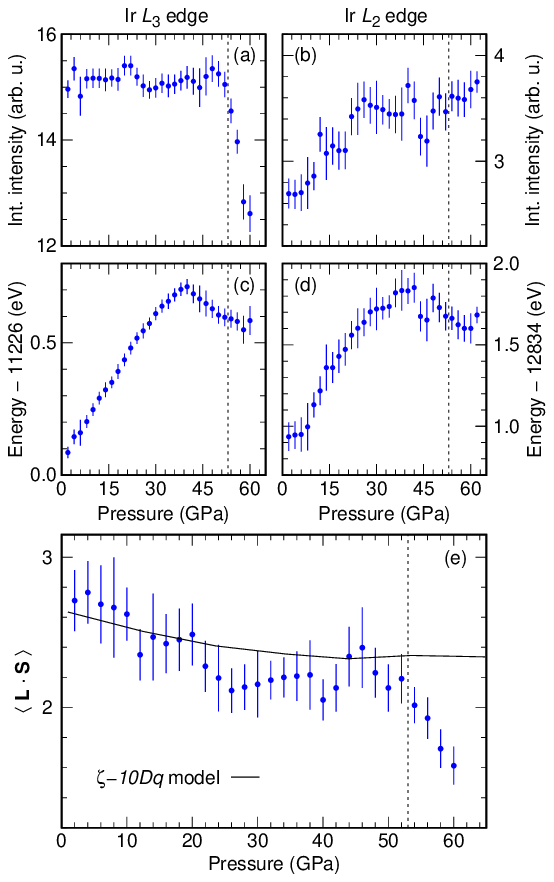}
\caption{Fitted high-pressure XAS data of Sr$_3$Ir$_2$O$_7$. (a, b) White line integrated intensities at the Ir $L_3$, $L_2$ edge. (c, d) White line energy position (absorption maximum) of the Ir $L_3$, $L_2$ edge. (e) Ground state SOC expectation value $\braket{{\bf L} \cdot {\bf S}}$, deduced from the branching ratio of integrated intensities of white lines, $I_{L_3}/I_{L_2}$. The solid line is the calculated $\braket{{\bf L} \cdot {\bf S}}$ of the $\zeta-10Dq$ model, using $\zeta=0.45$ eV and $10Dq$ values from Ref. \onlinecite{ding2016pressure}. In all panels, the vertical dotted line denotes the structural phase transition at 53 GPa.}
\label{fig2}
\end{figure}

However, as Laguna-Marco \emph{et al.} \cite{laguna-marco2010orbital} pointed out, the SOC expectation value exceeds that expected from a $J_{\text{eff}}=1/2$ state: If we consider the ground state as one hole in the $J_{\text{eff}}=1/2$ doublet and four holes in the $e_g$ levels, only the $J_{\text{eff}}=1/2$ state would contribute to the SOC expectation value, as $\braket{{\bf L} \cdot {\bf S}}_{e_g}=0$, and we obtain a total $\braket{{\bf L} \cdot {\bf S}} = 1$. Thus a model that considers SOC acting on the isolated $t_{2g}$ submanifold cannot accurately describe XAS data of iridates. Generally, this approximation is valid when the spin-orbit coupling $\zeta$ can be treated as a weak perturbation on the cubic CF $10Dq$, i.e. when $\zeta \ll 10Dq$. However, in Ir$^{4+}$ iridates, typically, $\zeta \sim \frac{1}{6} 10Dq$, which makes this assumption questionable. Indeed, using configuration interaction calculations, Laguna-Marco \emph{et al.} have shown that the SOC hybridizes $J_{\text{eff}}=3/2$ and $e_g$ levels \cite{laguna-marco2010orbital}. The hybridized $e_g$ levels then acquire non zero $\braket{{\bf L} \cdot {\bf S}}_{e_g}$, and SOC expectation values comparable to the ones derived from the BR could be obtained.

In order to accurately extract the BR from our data, we used a simple model that captures the salient features of the XAS spectra. Following Ref. \onlinecite{clancy2012spin}, we fitted the absorption step and white line with an arctangent and a Lorentzian function (see Fig. \ref{fig1} and Appendix \ref{appendixA}). The fitted center and width were identical for arctangent and Lorentzian functions. While the fit fails to reproduce details, such as oscillations above the absorption edge, it allows to accurately track the evolution of the white line in a robust manner. This was confirmed by numerical integration of the white line after subtracting the arctangent step, which, within error bars, yielded identical BRs (data not shown). Due to the large number of collected pressure points ($\sim$ 100 spectra per absorption edge), we binned the data into 2 GPa intervals.

Figure \ref{fig2} shows the results of the fitting procedure. Panels (a-b) show that up to the structural transition, the integrated intensity of the $L_3$ edge white line stays approximately constant, whereas an increase in intensity occurs at the $L_2$ edge. Across the structural transition ($\sim$ 53 GPa), the $L_3$ edge white line intensity shows a rapid decrease, while the $L_2$ edge intensity is largely unaffected. The resulting BR, converted to $\braket{{\bf L} \cdot {\bf S}}$ as described above, is plotted in panel (e). Panels (c, d) show the evolution of the energy position of the white line (defined as the absorption maximum) at $L_{3,\,2}$ edges. At both edges, the white line energy increases approximately linearly with pressure, followed by an anomaly incipient at $\sim 40$ GPa.

\subsection{Branching ratio in the low-pressure phase}
\label{LP-BR}

We first discuss the pressure evolution of the BR up to the structural transition. From ambient pressure to 50 GPa, the BR gradually decreases from $\sim 5.5$ to $\sim 4$. Rewriting $\braket{{\bf L} \cdot {\bf S}} = n_h {(\text{BR}-2)}/{(\text{BR}+1)}$, this corresponds to $\braket{{\bf L} \cdot {\bf S}}$ reducing from $\sim 2.7$ to $\sim 2.3$. 

\begin{figure}[t]
\includegraphics[width=\linewidth]{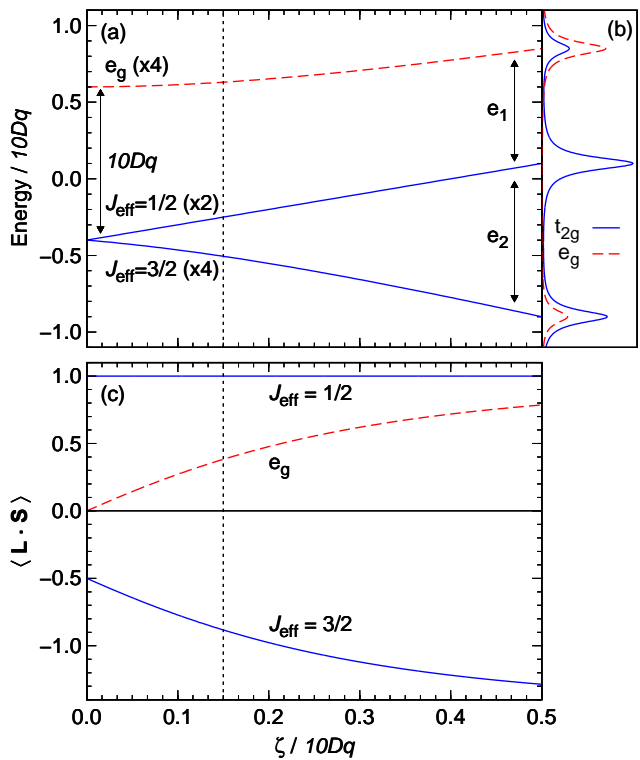}
\caption{Spin-orbit induced hybridization of $t_{2g}$ and $e_{g}$ levels within $\zeta-10Dq$ model. (a) $5d$ energy levels as a function of $\zeta/10Dq$. The energy differences $e_1$ and $e_2$ are defined in Appendix \ref{appendixB}. (b) Projection of $5d$ wavefunctions onto $t_{2g}$ (blue, solid line) and $e_g$ (red, dashed line) basis states at $\zeta/10Dq=0.5$. (c) $\braket{{\bf L} \cdot {\bf S}}$ contributions of individual $5d$ levels as a function of $\zeta/10Dq$. In panels (a) and (c) the vertical dotted line denotes a characteristic value of $\zeta/10Dq = 0.15$ for Ir$^{4+}$ iridates at ambient pressure.}
\label{fig3}
\end{figure}

Following the argument by Laguna-Marco \emph{et al.} \cite{laguna-marco2010orbital}, we here consider $t_{2g} - e_{g}$ hybridization within a single-ion model of SOC $\zeta$ and cubic CF splitting $10Dq$. While a single-ion approach fails to account for band effects, such as covalency and electron-electron interactions, it has the advantage of focusing on the two dominant energy scales and therefore requires only a single adjustable parameter, $\zeta/10Dq$. We diagonalized the Hamiltonian $\mathcal{H} = \mathcal{H}_{\text{SOC}} +\mathcal{H}_{\text{Cubic-CF}}$ in a complete basis of $d$-orbitals (see Appendix \ref{appendixB}). The resulting CF levels and corresponding $\braket{{\bf L} \cdot {\bf S}}$ values are plotted in Fig. \ref{fig3}, as a function of $\zeta/ 10Dq$. Effectively, this $\zeta-10Dq$ model allows to continuously tune the electronic state from a $J_{\text{eff}} = 1/2,\,3/2$ and $e_g$ splitting ($\zeta \ll 10Dq$) to a $J = 3/2,\,5/2$ splitting of the $d$ manifold ($\zeta \gg 10Dq$). As $\zeta/ 10Dq$ increases, $J_{\text{eff}} = 3/2$ and $e_g$ levels start to hybridize, which has a pronounced effect on their $\braket{{\bf L} \cdot {\bf S}}$ values. It is interesting to note that the orbital character of the $J_{\text{eff}} = 1/2$ doublet is unaffected. Hence models relying on a single hole in a $J_{\text{eff}} = 1/2$ state remain accurate even in the presence of SOC-induced $t_{2g}-e_g$ mixing. However, when other states are involved, as is the case in a XAS experiment, $t_{2g} - e_{g}$ hybridization should be taken into account.

Near ambient pressure, our XAS data of Sr$_3$Ir$_2$O$_7$ yields $\braket{{\bf L} \cdot {\bf S}} \sim 2.7$. For iridates, typical empirical values of $\zeta$ and $10Dq$ are 0.45 eV \cite{kim2012magnetic, liu2012testing, gretarsson2013crystal-field, moretti2014cairo} and 3 eV \cite{gretarsson2013crystal-field, moretti2014crystal, ding2016pressure}, respectively; thus $\zeta/ 10Dq = 0.15$. The ground state SOC expectation value can then be computed as $\braket{{\bf L} \cdot {\bf S}} = 4 \braket{{\bf L} \cdot {\bf S}}_{e_g}+\braket{{\bf L} \cdot {\bf S}}_{J_{\text{eff}}=1/2} \approx 2.5$. This is indeed close to the experimental estimate, and suggests that the $\zeta-10Dq$ model can accurately describe the observed BR. However, it must be emphasized that some variation exists in the value of the BR and the strength of the cubic CF reported in the literature \cite{haskel2012pressure, clancy2012spin, moretti2014crystal, ding2016pressure}. Without clear consensus in the experimental data, precise energy scales cannot be extracted. Nevertheless, the evolution of the BR within a consistent experimental setup offers valuable insights into the character of the electronic ground state.

As pressure is applied, the strength of the cubic CF increases, reducing $\zeta/10Dq$ and hence quenching $\braket{{\bf L} \cdot {\bf S}}_{e_g}$. Empirical values of $10Dq$ in Sr$_3$Ir$_2$O$_7$ as a function of pressure were determined by RIXS \cite{ding2016pressure}. Using these $10Dq$ values and a constant SOC of $\zeta=0.45$ eV \cite{kim2012magnetic, liu2012testing, gretarsson2013crystal-field, moretti2014cairo} (we assume that the $\zeta$ is mainly determined by the nuclear charge and therefore pressure independent), we computed $\braket{{\bf L} \cdot {\bf S}}$ within the $\zeta-10Dq$ model. Figure \ref{fig2}(e) shows that despite its simplicity, the model offers a good description of the XAS data, up to 53 GPa. The observed reduction in the BR up to the structural transition can thus be understood from an increase in $10Dq$ with pressure, which suppresses the SOC-induced $t_{2g}-e_g$ hybridization.

We also note that the sum of white line intensities at $L_2$ and $L_3$ edges increases by $\sim 5$\% up to the phase transition. Appealing to the absorption sum rule, this indicates that the number of holes in the Ir $5d$ states increases with pressure \cite{stohr2006magnetism}. This is supported by density functional theory (DFT) calculations, which suggest that the strong Ir-O hybridization at high-pressure increases the number of holes on the Ir site \cite{donnerer2016pressure}. However, at all pressures, taking into account a pressure-dependent number of holes has a negligible effect on the obtained $\braket{{\bf L} \cdot {\bf S}}$ values (see Appendix \ref{appendixC}).

\subsection{Pressure dependence of white line energy}

We now discuss the pressure dependence of the white line energy, defined as the absorption maximum. We observe a linear increase of 17.4(2) and 26(1) meV/$\,$GPa at the $L_3$ and $L_2$ edges, respectively (note that the absorption threshold, defined as the inflection point of the XAS, shows the same pressure dependence). The leading order term of a cubic CF is a spherically symmetric potential, inversely proportional to the Ir-O distance, which raises the energy of all $d$ levels. Assuming the $2p$ core levels are sufficiently screened from the CF, as a function of pressure, this term should approximately linearly increase the energy required to promote an electron from $2p$ to $5d$ levels \cite{donnerer2016pressure}.

The white-line energy increases more rapidly with pressure at the $L_2$ edge. Selection rules imply that only XAS transitions to empty $e_g$ levels are allowed at the $L_2$ edge, not to the half-occupied $J_{\text{eff}}=1/2$ doublet. The energy position of the $L_2$ edge white line will hence be determined by the energy of $e_g$ states. At the $L_3$ edge, transitions to both $J_{\text{eff}}=1/2$ and $e_g$ states are allowed, and the white line will be a sum of $t_{2g}$ and $e_g$ final states. This is reflected in the width of the $L_3$ edge white line, which is about 1 eV broader than the $L_2$ edge white line (individual $t_{2g}$ and $e_g$ features cannot be discerned). As $10Dq$ increases with pressure, the energy of $t_{2g}$ levels is lowered relative to $e_g$ states, and hence the overall $L_3$ edge energy will be raised less. In this case, we would also expect that the width of the white line at the $L_3$ edge increases more rapidly with pressure. However, within the experimental uncertainty, the increase in width is identical for both edges (about 9 meV/ GPa), indicating that it is dominated by bandwidth and core-hole lifetime broadening.

Above 40 GPa, the white line energy of both edges shifts to lower energies. This behavior could be related to the IMT: In a metallic state, additional transitions to the Fermi level become allowed, which could move the white line to lower energies. Nevertheless, this cannot explain the energy shift at the $L_2$ edge, where only transitions to $e_g$ states are allowed, and one should not be sensitive to the appearance of a $J_{\text{eff}}=1/2$ Fermi surface. Furthermore, the observed energy shift is substantially larger than the insulating gap ($\sim 100$ meV). We hence deem an IMT unlikely as the cause of this anomaly.

An alternative interpretation is that the strength of the cubic CF decreases above 40 GPa. This would explain the energy shift at both absorption edges, as well as the larger effect observed at the $L_2$ edge. Indeed, RIXS measurements indicate that $10Dq$ reaches a maximum at 40 GPa and then decreases at higher pressures \cite{ding2016pressure}. However, the microscopic origin of this effect is unclear.

\subsection{Branching ratio in the high-pressure phase}

Finally, we discuss the XAS data across the high-pressure structural transition of Sr$_3$Ir$_2$O$_7$. Above 50 GPa, we observe a further decrease in the BR, which originates from the $L_3$ edge white line intensity. At 60 GPa, the BR reaches $\sim$ 3, which corresponds to $\braket{ {\bf L} \cdot {\bf S}} \sim 1.6$. Within the $\zeta-10Dq$ model, the ratio of SOC to cubic CF would have to reach $\zeta / 10Dq \sim 0.05$ in order to yield this $\braket{ {\bf L} \cdot {\bf S}}$ value. This would require either $10Dq \sim 10$ eV at constant $\zeta = 0.45$ eV, or, alternatively, $\zeta \sim 0.18$ eV at $10Dq = 3.5$ eV. Both options appear unphysical and incompatible with high-pressure RIXS data \cite{ding2016pressure}, and we conclude that a single-ion model of SOC and cubic crystal fields cannot provide an adequate description of the electronic state of Sr$_3$Ir$_2$O$_7$ at high-pressure.

While uncertainties remain regarding the details of the high-pressure structure of Sr$_3$Ir$_2$O$_7$, the first-order nature of the transition and resulting monoclinic unit cell could result in significant changes to the local environment of the Ir ion \cite{donnerer2016pressure}. For example, if the transition results in strong non-cubic crystal fields, the $J_{\text{eff}} = 1/2$ state will not be fully realized, resulting in a suppression of $\braket{{\bf L} \cdot {\bf S}}$ \cite{moretti2014resonant}. Another interpretation of a reduced BR at high-pressure was provided in a XAS study of Sr$_2$IrO$_4$, where it was argued that a continuous, bandwidth-driven mixing of $J_{\text{eff}}=1/2,\,3/2$ levels occurs with pressure \cite{haskel2012pressure}. Most recently, a decrease in the BR was observed at only 2 GPa in $\beta$-Li$_2$IrO$_3$, which the authors attributed to a reduction in the effective electronic correlations \cite{veiga2017pressure}.

It is insightful to compare our observations to high-pressure RIXS data of Sr$_3$Ir$_2$O$_7$, where the $\ket{J_{\text{eff}}=1/2} \rightarrow \ket{J_{\text{eff}}=3/2}$ crystal field excitation was measured \cite{ding2016pressure}. This feature is sensitive to the amount of CF distortions and bandwidth of $J_{\text{eff}}$ states \cite{ament2011theory}. Above the structural transition, the excitation energy of this feature increased by $15\%$ compared to its ambient pressure value \cite{ding2016pressure}. This increase in excitation energy could suggest a splitting of $J_{\text{eff}}=3/2$ states through non-cubic CFs, which may not be resolvable due to large $t_{2g}$ bandwidths. 

While the our data shows that a reconstruction of the $5d$ states occurs at the high-pressure transition of Sr$_3$Ir$_2$O$_7$, the resulting electronic state remains uncertain. Nevertheless, we can make the following conclusions. First, while the BR at 60 GPa is substantially reduced from its ambient pressure value, the resulting $\braket{{\bf L} \cdot {\bf S}}$ value is not yet fully quenched, which implies that the spin-orbit coupling still affects the electronic ground state. Whether the remaining $\braket{{\bf L} \cdot {\bf S}}$ originates solely from holes in the $e_g$ levels, or still has a considerable contribution from the hole in the $J_{\text{eff}}=1/2$ level, cannot be ascertained from our data. Second, we note that all of the above scenarios leading to a reduced BR will affect the orbital character of the hole in the ground state doublet. In particular, this suggests that in the high-pressure phase of Sr$_3$Ir$_2$O$_7$ a significant departure from the ideal $J_{\text{eff}}=1/2$ limit occurs.
 
\section{Conclusion}

We have performed high-pressure XAS at the Ir $L_3$ and $L_2$ absorption edges of Sr$_3$Ir$_2$O$_7$. The branching ratio of white-line intensities allowed us to assess the character of the electronic ground state. Up to the structural transition at 53 GPa, the BR gradually reduces from $\sim 5.5$ to $\sim 4$ (corresponding to $\braket{ {\bf L} \cdot {\bf S}} \sim 2.7$ and $\sim 2.3$, respectively). This could be understood within a single-ion model that takes into account the spin-orbit induced hybridization of $t_{2g}$ and $e_{g}$ levels. As the cubic CF increases with pressure, SOC-induced mixing of $J_{\text{eff}}=3/2$ and $e_g$ states is suppressed, reducing the $\braket{ {\bf L} \cdot {\bf S}}$ contribution of $e_g$ levels. Above the structural transition, the BR decreases further, eventually yielding $\braket{ {\bf L} \cdot {\bf S}} \sim 1.6$ (BR $\sim$ 3) at 60 GPa. This could be driven by non-cubic crystal fields or bandwidth-driven mixing of $J_{\text{eff}}=1/2,\,3/2$ levels, and indicates that the high-pressure electronic state of Sr$_3$Ir$_2$O$_7$ significantly deviates from the $J_{\text{eff}}=1/2$ state known at ambient conditions.

Data presented in this paper can be obtained from Ref. \cite{dataset}.

\begin{acknowledgments}
We would like to thank J. Jacobs for the preparation and gas-loading of the DACs. We acknowledge support from R. Torchio, S. Boccato, and O. Mathon. We acknowledge useful discussions with D. Haskel. This work is supported by the UK Engineering and Physical Sciences Research Council under Grants No. EP/J016713/1 and No. EP/N027671/1.

\end{acknowledgments}

\appendix

\section{Fitting of XAS data}
\label{appendixA}

We fitted the XAS data using

\begin{equation*}
\mu(E) = C_0 + C_1E + C_2 \arctan \left( \frac{E-E_0}{\Gamma/2} \right) + \frac{C_3}{1+\left( \frac{E-E_0}{\Gamma/2} \right)^2}
\end{equation*}
where $C_0$ and $C_1$ describe a first-order polynomial background, $C_2$ is the absorption step height, $C_3$ is the white line intensity and $E_0$ and $\Gamma$ are the center and width of both arctangent and Lorentzian functions. Following the fit, the polynomial background was subtracted from the data, and the data and fitted parameters were normalized to the respective absorption steps, as shown in Fig. \ref{fig1}. The white line integrated intensity is given by $I_{L_{2,3}} = C_3 \Gamma$.

\section{$\zeta-10Dq$ model}
\label{appendixB}

The Hamiltonian of spin-orbit coupling $\zeta$ and cubic CF $10Dq$, in a basis of $e_g$ and $t_{2g}$ orbitals, can be written as

\begin{align*}
\mathcal{H} &= \mathcal{H}_{\text{SOC}} + \mathcal{H}_{\text{Cubic-CF}}\\
            &= \zeta {\bf L} \cdot {\bf S} + Dq \left( 6 d^\dagger_{e_g,\sigma} d_{e_g,\sigma} - 4 d^\dagger_{t_{2g},\sigma} d_{t_{2g},\sigma} \right).
\end{align*}

\noindent We will not show the explicit eigenfunctions for lack of space. Their energies are\\

\begin{table}[!htb]
\centering
\begin{tabular}{l c l}
$E_{{e_g}} $ & $=$ & $\frac{1}{20} \left( 2 \Delta - 5 \zeta + 5 c \right)$,\\
\rule{0pt}{1ex} \\
$E_{{J_{\text{eff}}=1/2}} $ & $=$ & $\frac{1}{5} \left( -2 \Delta + 5 \zeta \right)$,\\
\rule{0pt}{1ex} \\
$E_{{J_{\text{eff}}=3/2}} $ & $=$& $\frac{1}{20} \left( 2 \Delta - 5 \zeta - 5 c \right)$,\\
\end{tabular}
\end{table}

\noindent where $c = \sqrt{4\Delta^2 + 4 \Delta \zeta + 25 \zeta^2}$ and we have defined $\Delta = 10 Dq$ for ease of notation. For simplicity, we have kept the labels $\{ \ket{e_g}, \ket{J_{\text{eff}}=1/2},\ket{J_{\text{eff}}=3/2} \}$, which strictly apply only in the absence of $t_{2g}-e_g$ hybridization. 

The resulting intra-$t_{2g}$ excitation energies are

\begin{table}[!htb]
\centering
\begin{tabular}{ r l l }
 $e_1=$ & $\frac{1}{4} \left( 2 \Delta - 5 \zeta + c \right)$ & for $\ket{J_{\text{eff}}=1/2} \rightarrow \ket{e_g}$, \\
 \rule{0pt}{1ex} \\
 $e_2=$ & $\frac{1}{4} \left( -2 \Delta + 5 \zeta + c \right)$ & for $\ket{J_{\text{eff}}=1/2} \rightarrow \ket{J_{\text{eff}}=3/2}$, \\
\end{tabular}
\end{table}

\noindent The expectation values of the spin-orbit coupling are

\begin{table}[!htb]
\centering
\begin{tabular}{ l c l }
 $\braket{{\bf L} \cdot {\bf S}}_{{e_g}}$ & $=$ & $\frac{1}{4c} \left( 2 \Delta + 25 \zeta - c \right)$, \\
  \rule{0pt}{1ex} \\
 $\braket{{\bf L} \cdot {\bf S}}_{{J_{\text{eff}}=1/2}}$ & $=$ & $1$, \\
  \rule{0pt}{1ex} \\
 $\braket{{\bf L} \cdot {\bf S}}_{{J_{\text{eff}}=3/2}}$ & $=$ & $-\frac{1}{4c} \left( 2 \Delta + 25 \zeta + c \right)$. \\ 
\end{tabular}
\end{table}


\section{Pressure dependence of number of holes in the Ir $5d$ band}
\label{appendixC}

In Section \ref{LP-BR}, we observed that the sum of $L_3$ and $L_2$ white line intensities of Sr$_3$Ir$_2$O$_7$ is not constant with pressure. As our measurements were performed on powder samples, the combined intensity of $L_3$ and $L_2$ resonances should be proportional to the number of holes in the Ir $5d$ states, $n_h$ \cite{stohr2006magnetism}.

\begin{figure}[!htb]
\includegraphics[width=\linewidth]{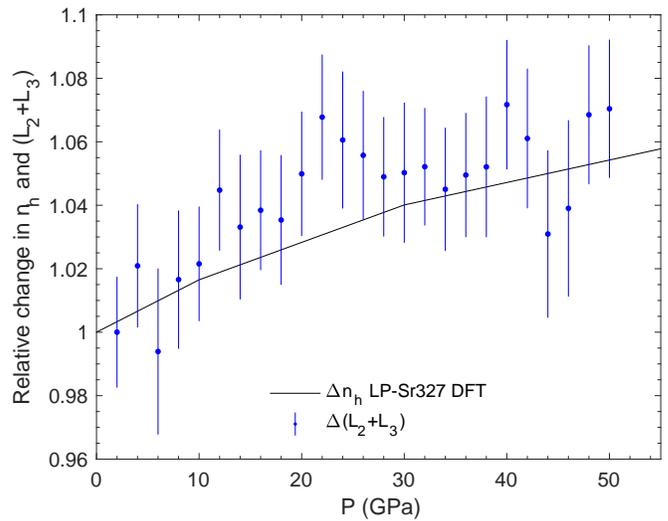}
\caption{Relative change in the number of holes of Sr$_3$Ir$_2$O$_7$. The blue circles are the relative change in the sum of $L_3+L_2$ intensities, the black line is the relative change in $n_h$, as calculated by DFT in Ref.  \cite{donnerer2016pressure}.}
\label{fig1s}
\end{figure}

In the low-pressure phase of Sr$_3$Ir$_2$O$_7$, the sum of $L_3$ and $L_2$ white line intensities increases by $\sim 5$\% up to the phase transition, indicating an increase in $n_h$. DFT calculations of Ref. \cite{donnerer2016pressure} have proposed that strong Ir-O hybridization at high-pressure increases the number of holes on the Ir site. Figure \ref{fig1s} shows that the relative change in the sum of $L_3$ and $L_2$ white line intensities agrees well with the relative change in $n_h$ calculated by DFT, suggesting that Ir-O hybridization is responsible for increasing $n_h$ in the low-pressure phase Sr$_3$Ir$_2$O$_7$.

\begin{figure}[!htb]
\includegraphics[width=\linewidth]{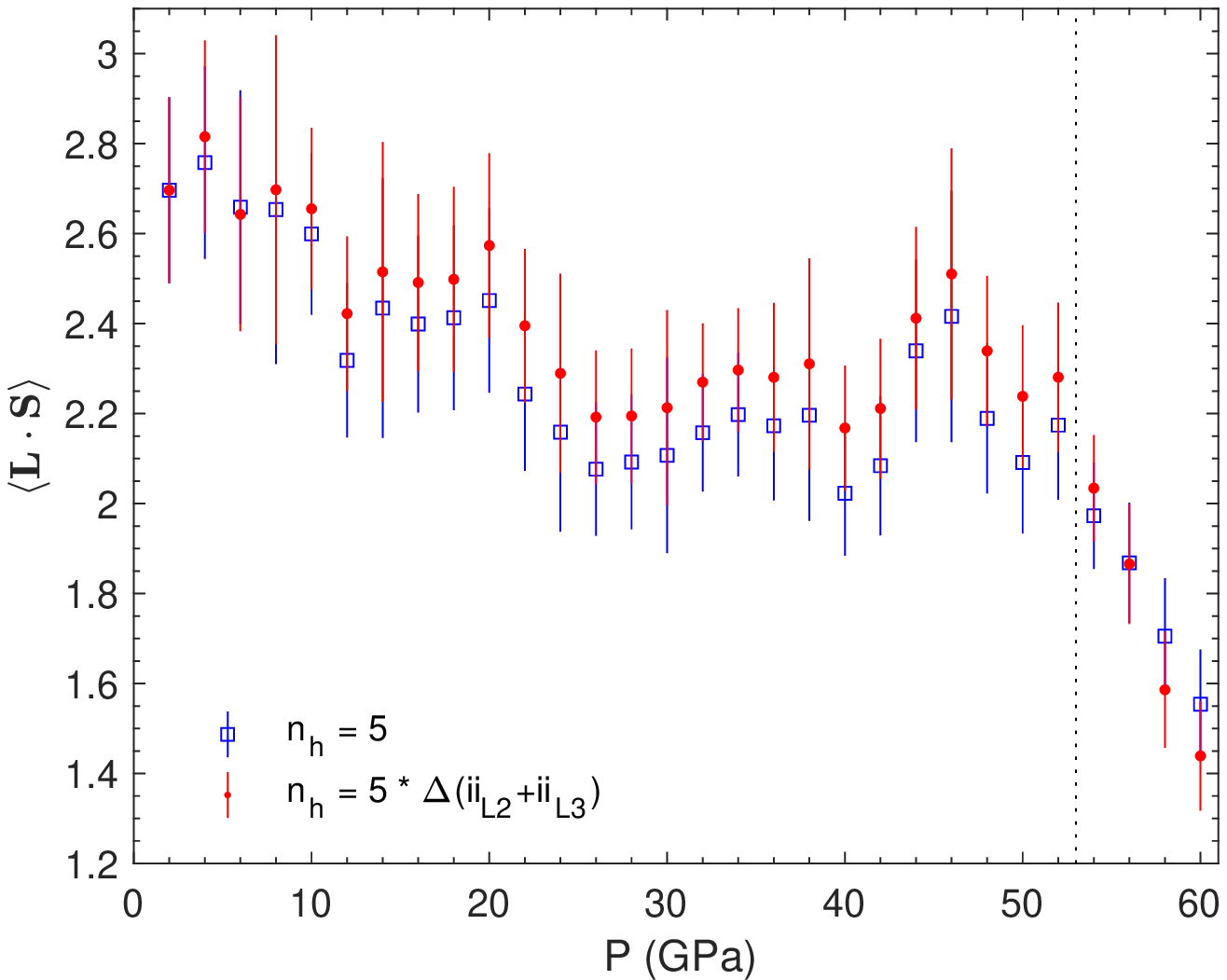}
\caption{Effect of pressure-dependent number of Ir $5d$ holes ($n_h$) on the ground state expectation value $\braket{{\bf L} \cdot {\bf S}}$ of Sr$_3$Ir$_2$O$_7$. The blue squares are $\braket{{\bf L} \cdot {\bf S}}$ values obtained with a constant $n_h=5$; the red circles are $\braket{{\bf L} \cdot {\bf S}}$ values determined with a pressure-dependent $n_h$, estimated by scaling an ambient $n_h=5$ to the change in the sum of $L_3+L_2$ intensities.}
\label{fig2s}
\end{figure}

In our data analysis, we have assumed a constant $n_h = 5$ to convert the branching ratio to $\braket{{\bf L} \cdot {\bf S}}$, via $\braket{{\bf L} \cdot {\bf S}} = n_h {(\text{BR}-2)}/{(\text{BR}+1)}$. We estimate how a pressure-dependent $n_h$ would affect the obtained $\braket{{\bf L} \cdot {\bf S}}$ value, by scaling $n_h$ from an ambient value of $n_h=5$ with the change in the sum $L_3+L_2$ white line intensities. Figure \ref{fig2s} shows how $\braket{{\bf L} \cdot {\bf S}}$ would differ in this case. We conclude that, within the uncertainty of the fitted data, introducing a pressure-dependent $n_h$ has a negligible effect on the obtained $\braket{{\bf L} \cdot {\bf S}}$ values.

\end{document}